\documentclass[sigconf,nonacm,natbib=false]{acmart}

\AtBeginDocument{%
  }

\usepackage{iftex}
\usepackage{latexml}
\usepackage{xcolor}
\definecolor{zhtranslationcolor}{rgb}{0.16,0.22,0.48}

\usepackage{booktabs}   
\usepackage{array}      %

\newcommand{\li}{\textcolor{black}}
\usepackage{graphicx}   

\usepackage{amsmath}
\newcommand{\ms}[2]{{#1\tiny{$\pm$#2}}}
\usepackage{stfloats}   %
\usepackage{pifont}

\usepackage{booktabs}   
\usepackage{graphicx}   
\usepackage{array}      %
\usepackage{booktabs}   
\usepackage{graphicx}   
\usepackage{amsmath}    %
\usepackage[normalem]{ulem}       

\newif\ifshowrevisioncolors
\showrevisioncolorsfalse
\ifshowrevisioncolors
  \newcommand{\addtext}[1]{\textcolor{red}{#1}}
  \newcommand{\addtextblue}[1]{\textcolor{blue}{#1}}
\else
  \newcommand{\addtext}[1]{\textcolor{black}{#1}}
  \newcommand{\addtextblue}[1]{\textcolor{black}{#1}}
\fi

\usepackage{amsmath}
\usepackage{algorithm}
\usepackage{algorithmic}

\usepackage{multirow}
\usepackage{placeins}

\usepackage{xurl}        %
\usepackage{hyperref}
\definecolor{linkblue}{rgb}{0.05,0.25,0.60}
\hypersetup{
  breaklinks=true,
  colorlinks=true,
  linkcolor=linkblue,
  citecolor=linkblue,
  urlcolor=linkblue,
  filecolor=linkblue
}
\AtEndPreamble{%
  \hypersetup{
    colorlinks=true,
    linkcolor=linkblue,
    citecolor=linkblue,
    urlcolor=linkblue,
    filecolor=linkblue
  }%
  \urlstyle{tt}%
}

\usepackage[most]{tcolorbox}
\usepackage{xcolor}

\definecolor{promptbgcolor}{RGB}{245,245,245} %
\definecolor{promptframecolor}{RGB}{160,160,160} %

\newtcolorbox{promptbox}[1]{
  colback=promptbgcolor,
  colframe=promptframecolor,
  coltitle=black,
  title={\textbf{#1}}, %
  fonttitle=\small\sffamily,
  fontupper=\small\ttfamily, %
  boxrule=0.8pt,
  arc=2pt,
  left=5pt, right=5pt, top=5pt, bottom=5pt,
  breakable %
}

\setcopyright{none}
\makeatletter
\renewcommand{\footnotetextcopyrightpermission}[1]{}
\makeatother

\RequirePackage[
  datamodel=acmdatamodel,
  style=acmnumeric,
  doi=false,
  url=false,
  minbibnames=9,
  ]{biblatex}

\begin{document}
\raggedbottom

\title{Agent4Crypto: A Dual-Reflective and Market-Feedback-Aligned LLM-Based Multi-Agent System for Cryptocurrency Trading}

\author{Guangyan Gan}
\authornote{Equal contribution.}
\affiliation{%
  \institution{Nanyang Technological University}
  \city{Singapore}
  \country{Singapore}
}
\email{GUANGYAN001@e.ntu.edu.sg}

\author{Jiahao Li}
\ifdefined\iflatexml\iflatexml\authornote{Equal contribution.}\fi\fi
\authornotemark[1]
\affiliation{%
  \institution{Tsinghua University}
  \city{Beijing}
  \country{China}
}
\email{lijh@pbcsf.tsinghua.edu.cn}

\author{Jiashun Wang}
\affiliation{%
  \institution{Shanghai University of International\\ Business and Economics}
  \city{Shanghai}
  \country{China}
}
\email{24089054@suibe.edu.cn}

\author{Lin William Cong}
\authornote{Corresponding author.}
\affiliation{%
  \institution{Nanyang Technological University}
  \city{Singapore}
  \country{Singapore}
}
\email{will.cong@ntu.edu.sg}

\renewcommand{\shortauthors}{Gan, Li, Wang, and Cong}

\begin{abstract}
Cryptocurrency markets provide a natural testbed for autonomous AI agents because they are digital-native, high-frequency, and machine-speed environments, yet existing methods often struggle with non-stationary market regimes, opaque decisions, and fixed aggregation rules. In this paper, we propose Agent4Crypto, a dual-reflective and market-feedback-aligned multi-agent system. The framework coordinates specialized agents for data collection, analysis, risk assessment, reflection, and trading, synthesizing heterogeneous signals from prices, indicators, news, and charts in an interpretable decision pipeline. Our central design is a trader-centric decision mechanism: three style-specific trader agents, Aggressive, Neutral, and Conservative, generate candidate actions and natural-language rationales from the same market context, while their influence is learned rather than manually fixed. A PPO-based gating policy dynamically allocates trader weights through a dual-reflective feedback process. One reflection stream evaluates post-trade decision quality from realized portfolio outcomes and distills reusable lessons into a Financial ReasoningBank for retrieval in subsequent trader prompts, while the other estimates trader-level counterfactual contribution, indicating which style has recently aligned better with market movements. The PPO state combines Counterfactual Contribution Memory with market features, previous trader weights, and current trader actions, so trader influence adapts across bull, sideways, and bear regimes. Experiments on BTC, ETH, and XRP show that Agent4Crypto improves total return and Sharpe ratio over traditional strategies, machine learning and reinforcement learning baselines, and existing LLM-based trading systems, demonstrating the value of interpretable trader competition and market-feedback-driven adaptation in volatile cryptocurrency markets.
\end{abstract}



\maketitle

\section{Introduction}  
Cryptocurrency trading presents unique challenges due to extreme market volatility, speculative behavior, market inefficiency, and the absence of well-established asset pricing theories~\cite{li2023online,Fang2022CryptocurrencySurvey}. Unlike traditional financial markets, cryptocurrencies represent the first AI-native financial environments, where machine-first incentives, machine-speed interactions, and programmable rules naturally enable autonomous agents to participate and learn. Specifically, crypto markets operate 24/7, provide standardized exchange APIs for machine-speed execution, and expose publicly observable on-chain and news signals that can be integrated into automated decision pipelines. In contrast to conventional markets, where human decisions dominate and AI adoption is gradual, crypto markets already compel AI systems to compete, adapt, and evolve at scale. This makes them ideal laboratories for designing, testing, and refining AI-driven trading strategies before similar approaches become feasible in traditional finance.  

However, these characteristics also result in frequent and abrupt market regime shifts, under which trading strategies that perform well in one period may quickly degrade. Effectively adapting trading decisions to such regime changes remains a central challenge in cryptocurrency markets.
Developing robust trading strategies in this environment requires integrating heterogeneous information sources, including price dynamics, technical indicators, market sentiment, and external events, under highly stochastic and non-stationary conditions. In practice, this process demands substantial domain expertise and intensive manual effort, placing a significant burden on financial professionals.

A growing body of research has explored machine learning (ML) and deep learning (DL)-based approaches for financial trading~\cite{Goutte2023DeepMarket,Lahmiri2019CryptocurrencyNetworks,li2026sapp}. Although these methods achieve promising predictive performance, most operate as black-box models with limited interpretability, raising concerns regarding trust, risk control, and robustness in real-world trading environments~\cite{Li2023PEN:Explainability}. These limitations are particularly pronounced in cryptocurrency markets, where regime shifts and extreme price movements amplify the consequences of poorly understood decisions.

Recent advances in large language models (LLMs) introduce a new paradigm for financial decision-making. LLMs demonstrate strong capabilities in reasoning over heterogeneous and multimodal information, including textual data such as news and reports~\cite{Li2023LargeSurvey} and visual signals such as price charts~\cite{Zhang2024Vision-LanguageSurvey}. Moreover, their natural language generation abilities enable more transparent and interpretable trading decisions through explicit reasoning processes~\cite{wei2022chain}. Building on these strengths, several LLM-based financial agent systems, such as \textsc{FinGPT}~\cite{yang2023fingpt} and \textsc{FinMem}~\cite{yu2023finmem}, have shown encouraging results. However, when deployed as single-agent systems, LLMs often exhibit limited performance gains in trading due to the complexity of financial decision-making and the scarcity of domain-specific knowledge in highly volatile cryptocurrency markets~\cite{Li2023MultimodalResearch}.

To overcome the limitations of single-agent designs, recent studies increasingly adopt multi-agent LLM frameworks that decompose complex trading tasks into specialized subtasks handled by multiple agents~\cite{wu2024autogen}. Such systems enable collaboration among agents with complementary expertise, leading to more comprehensive reasoning and improved decision robustness. While multi-agent LLM architectures have demonstrated success in stock trading~\cite{xiao2024tradingagents,zhang2024multimodal}, their application to cryptocurrency trading remains underexplored. Existing approaches, such as CryptoTrade~\cite{li2024cryptotrade}, rely on out-of-the-box LLMs with static agent roles and fixed interaction schemes, lack explicit risk management modules and memory mechanisms, do not account for heterogeneous trader decision styles, and do not incorporate strategies to align trader behavior with real market feedback.

Aligning LLM-based trading systems with financial objectives poses additional challenges. Parameter-efficient fine-tuning methods, such as LoRA~\cite{hu2021lora} and QLoRA~\cite{dettmers2023qlora}, have been applied to adapt LLMs for financial tasks, but predominantly in single-agent settings~\cite{yang2023fingpt,zhao2024aligning}. Reinforcement learning from human feedback (RLHF) offers another alignment strategy. However, applying RLHF to multi-agent cryptocurrency trading is non-trivial. Human evaluators cannot reliably assess decision quality in highly stochastic markets, and designing reward functions that accurately capture long-term market dynamics is fundamentally challenging. Moreover, in multi-agent systems, determining which agents to train and how to allocate training is itself a complex problem.

To address these limitations, we introduce Agent4Crypto, an explainable, multimodal, multi-agent framework for cryptocurrency trading explicitly aligned with market feedback. Agent4Crypto organizes specialized agents into collaborative teams that integrate heterogeneous market information, perform risk-aware analysis, and reflect on past decisions through shared memory and feedback mechanisms. We propose both intra- and inter-team collaboration and competition. In intra-team collaboration, agents in the \textit{Data Team} jointly analyze heterogeneous inputs, including asset, technical, news, and visual signals. In inter-team collaboration, the \textit{Data Team}, \textit{Reflector}, \textit{Risk Manager}, and \textit{Trader Team} jointly refine decisions.

Within the \textit{Trader Team}, we instantiate heterogeneous trader agents corresponding to distinct trading styles, including aggressive, neutral, and conservative agents, each tailored to different market regimes. From a modeling perspective, this system can be viewed as a Mixture-of-Experts, where each trader agent specializes in a particular regime. While such decomposition enables collaboration among agents with complementary expertise, existing LLM-based trading systems typically rely on static roles or heuristic aggregation schemes, which fail to adapt agent influence under rapidly changing market conditions.
We address this by formulating the aggregation of trader agents as a simplex-constrained, Mixture-of-Experts-style gating problem. \addtext{Instead of fine-tuning LLM parameters or directly optimizing individual trader policies, we introduce a two-level market-feedback memory design. First, a Financial ReasoningBank distills completed trading trajectories into interpretable success and failure lessons that can be retrieved for future trader reasoning. Second, a Counterfactual Contribution Memory (CCM) replays each trader's standalone recommendation against the next realized market move, producing trader-level contribution scores that are fed into the gating policy.} This policy is optimized via reinforcement learning, allowing adaptive reallocation of expert influence across regimes while preserving the interpretability of each trader’s behavior. In other words, \addtext{we propose a contribution-aware market-feedback alignment approach: rather than merely rewarding the final aggregated action, the system explicitly estimates which trader would have been correct under the realized market outcome and uses this counterfactual evidence to guide trust allocation.} Trader weights are dynamically updated through reinforcement learning, enabling the system to continuously adapt to evolving market conditions while maintaining transparent, interpretable decision-making.

This formulation enables the system to continuously adapt to evolving market conditions while preserving interpretability. It establishes a principled bridge between LLM-based reasoning, which excels at interpreting heterogeneous and multimodal market signals, and reinforcement learning, which provides a market-feedback-driven mechanism for adaptive aggregation of trading expertise.

Our main contributions are summarized as follows:
\begin{itemize}
    \item \addtext{We propose, to our knowledge, the first multi-agent cryptocurrency trading framework that combines interpretable financial experience memory with counterfactual trader-level contribution memory. The Financial ReasoningBank converts realized trading trajectories into retrievable success and failure lessons, while CCM estimates each trader's standalone hypothetical contribution from market feedback.}
    \item We design a unified collaboration and competition mechanism that coordinates heterogeneous agents and dynamically balances their influence in trading decisions while maintaining interpretability. Specifically, intra-team agents in the \textit{Data Team} collaboratively analyze multimodal information, inter-team agents (\textit{Data Team}, \textit{Reflector}, \textit{Risk Manager}, \textit{Trader Team}) collaborate on risk-aware and reflective decision-making, and \addtext{trader agents in the \textit{Trader Team} with distinct styles compete through a contribution-aware reinforcement learning (RL) gating policy that observes both current trader actions and their counterfactual contribution scores from recent history.}
    \item Extensive experiments \addtextblue{on BTC, ETH, and XRP} covering bullish, sideways/neutral, and bearish conditions show that Agent4Crypto exceeds the best evaluated baseline on all three assets.
\end{itemize}

 \section{Related Work}
In this section, we review recent advances in LLM-based trading systems and  aligning LLMs for financial applications.

\subsection{LLMs for Trading}  
Recent research has explored the use of LLMs as trading agents capable of generating investment decisions by analyzing diverse external signals, such as news, financial reports, and market prices \cite{ding2024large}. Early approaches primarily adopt a single-agent paradigm, where an LLM functions as an independent trader guided by prompt-based inputs. Examples include news-driven agents that infer market trends from financial news and sentiment analysis \cite{lopezlira2023chatgptforecaststockprice}. Other approaches incorporate reasoning and memory mechanisms to improve decision consistency, such as FinMem \cite{yu2023finmem}.  

While these single-agent methods demonstrate promising results, they often suffer from limited robustness and generalization due to their reliance on fixed reasoning patterns and the absence of diverse perspectives \cite{yan2025beyond}. To address these limitations, recent work has proposed multi-agent LLM-based trading frameworks, in which multiple agents with heterogeneous roles, such as news analysts, technical analysts, and risk controllers, interact through debate, collaboration, or information sharing \cite{xiao2024tradingagents,li2024cryptotrade,zhang2024multimodal,yu2024fincon,wang2024quantagent,xiong2025quantagent,singhi2025adaptive}. By encouraging diverse reasoning processes and cross-agent verification, these frameworks improve factual accuracy and produce more stable trading decisions.  

Despite these advances, existing studies rarely account for differences in trading style, including aggressive, neutral, or conservative styles, which play a critical role in real-world portfolio management. Furthermore, most multi-agent frameworks remain largely static: agent roles, interaction protocols, and relative contributions are manually designed and fixed during inference. In addition, these agents are typically neither fine-tuned nor explicitly aligned with actual market outcomes, and they do not learn agent contribution weights directly from realized trading returns, or profit and loss (P\&L), which limits their adaptability to evolving market conditions \cite{liu2025survey}.  

In contrast, our work advances LLM-based trading by introducing a market-aligned multi-agent framework that explicitly models heterogeneous trading styles and leverages real market feedback to adaptively adjust each agent's contribution. By employing reinforcement learning to dynamically weight agent decisions based on observed market outcomes, our framework achieves enhanced robustness, adaptability, and overall trading performance in complex cryptocurrency markets.

\subsection{Aligning LLMs for Financial Tasks}  
Effectively aligning LLM-based autonomous agents for finance-specific tasks involves multiple strategies. One common approach is retrieval-augmented generation \cite{gao2024optimized}, which incorporates external knowledge to improve response accuracy. Fine-tuning on domain-specific datasets \cite{wang2025parameter} further helps LLMs capture financial terminology and market nuances. Pretraining on large-scale financial corpora, as exemplified by BloombergGPT \cite{wu2023bloomberggpt} and BBT-Fin \cite{lu2023bbtfin}, improves domain alignment while preserving general language understanding.  

To reduce computational cost while maintaining adaptability, parameter-efficient fine-tuning methods such as LoRA \cite{hu2021lora} and QLoRA \cite{dettmers2023qlora} have been widely applied. Representative finance-oriented models, including PIXIU \cite{xie2023pixiu}, FinGPT \cite{yang2023fingpt}, and Instruct-FinGPT \cite{zhang2023instructfingpt}, leverage instruction-tuned financial datasets and consistently outperform base LLMs on classification and decision-making tasks in finance.  

Beyond static fine-tuning, feedback mechanisms provide an additional avenue for improvement. Self-refinement allows LLMs to iteratively evaluate and revise their outputs \cite{madaan2024self,wang2024quantagent}, while external feedback, such as reinforcement learning from human feedback \cite{ouyang2022training}, aligns model outputs with desired objectives. In financial contexts, reinforcement learning using market feedback has been applied to tasks such as sentiment prediction, enabling LLMs to dynamically adapt reasoning and decision-making policies. For instance, \cite{zhao2024aligning} demonstrates that updating LLM knowledge based on actual market outcomes can enhance predictive accuracy for future price movements.  

However, to the best of our knowledge, no prior work has integrated market feedback within a multi-agent LLM framework for trading. Our study addresses this gap by combining heterogeneous agents, diverse trading styles, and reinforcement learning-based market alignment, providing a unified and adaptive framework for robust cryptocurrency trading.

\section{Agent4Crypto Framework}
\label{sec:cryptoagent}

We present Agent4Crypto, a multi-agent framework designed for cryptocurrency trading. The framework aims to produce robust, adaptive, and interpretable trading decisions by coordinating multiple specialized agents aligned with market feedback. In addition to generating actionable trades, Agent4Crypto emphasizes transparent reasoning, explaining the rationale behind each decision.

\begin{figure*}[htbp]
\centering
\includegraphics[width=0.99\linewidth]{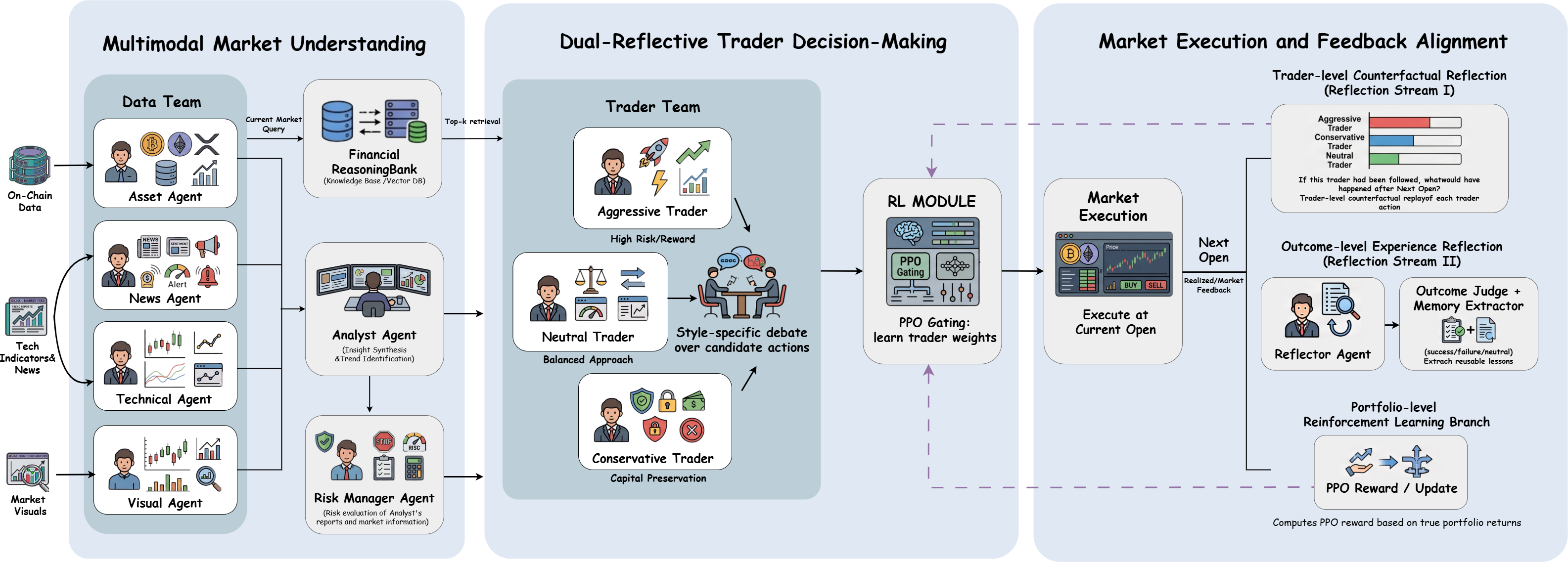}
\Description{A three-stage system diagram of Agent4Crypto. The first stage integrates multimodal market signals with a Data Team, Analyst Agent, Risk Manager Agent, and Financial ReasoningBank. The second stage uses three style-specific traders and PPO gating for weighted decision-making. The third stage executes trades and sends portfolio-level and trader-level feedback to reflection, memory, and reinforcement learning modules.}
\caption{\addtextblue{Three-stage architecture of Agent4Crypto. The Multimodal Market Understanding stage combines the Data Team, Analyst Agent, Risk Manager Agent, and Financial ReasoningBank. The Dual-Reflective Trader Decision-Making stage uses three style-specific traders and PPO gating to form a weighted action. The Market Execution and Feedback Alignment stage executes at the current Open and sends next-Open portfolio outcomes, trader-level counterfactual feedback, and distilled experience lessons back to the gating policy and memory.}}
\label{fig:architecture}
\end{figure*}

\subsection{Architecture Overview}

\addtextblue{Agent4Crypto follows a three-stage closed loop (Figure~\ref{fig:architecture}). First, the Multimodal Market Understanding stage uses the Asset, Technical, News, and Visual agents in the \textit{Data Team} to process heterogeneous signals. The \textit{Analyst Agent} synthesizes these signals, while the \textit{Risk Manager Agent} provides a qualitative risk assessment. The Financial ReasoningBank retrieves relevant prior lessons for the current market query. Second, the Dual-Reflective Trader Decision-Making stage lets the Aggressive, Neutral, and Conservative traders debate the same context. The Proximal Policy Optimization (PPO) Gating module learns their weights and forms the final continuous action. Third, the Market Execution and Feedback Alignment stage executes the action at the current Open and observes the outcome at the next Open. Outcome-level reflection distills lessons for memory, trader-level counterfactual reflection estimates each style's standalone contribution, and PPO updates the gating policy from portfolio reward and Counterfactual Contribution Memory (CCM). This organization connects multimodal understanding, interpretable trader competition, and market-feedback-driven adaptation.}

To enable adaptive and robust trading in dynamic markets, agents operate collaboratively and competitively, both within and across teams. These mechanisms are detailed in the next subsection.

\subsubsection{Collaboration and Competition}

Its agents leverage intra-team collaboration, inter-team collaboration, and intra-team competition to enhance decision quality and adaptability:

\begin{itemize}
    \item \textbf{Intra-team collaboration:} Within the \textit{Data Team}, specialized agents process heterogeneous inputs, including asset data, technical indicators, news, and visual charts, and jointly construct a comprehensive market view. This ensures downstream agents operate on unified, multimodal information.
    \item \textbf{Inter-team collaboration:} The \textit{Data Team}, \textit{Reflector}, \textit{Risk Manager}, and \textit{Trader Team} collaborate to refine trading decisions. Insights from historical reflection, risk assessment, and multimodal analysis are aggregated to guide the \textit{Trader Team} in executing profitable yet risk-aware actions.
    \item \textbf{Competition among traders:} Within the \textit{Trader Team}, agents representing distinct trading styles (aggressive, neutral, conservative) evaluate the same market context differently. Their outputs are combined using an RL-based gating mechanism, dynamically adjusting each agent's influence based on realized market performance.
\end{itemize}
This hierarchical collaboration and competition design ensures agents leverage complementary expertise, adapt to evolving market conditions, and maintain interpretable, robust trading behavior.

\subsection{Agent4Crypto Components}

\subsubsection{Data Team}
\label{datateam}

The \textit{Data Team} is responsible for acquiring, processing, and integrating diverse market data. It comprises four specialized agents, each focusing on a distinct aspect of market intelligence:

\begin{enumerate}
    \item \textit{Asset Agent.} \addtextblue{Uses asset-level on-chain and fundamental observations, such as transaction count, active addresses, native fees, and estimated market capitalization.} It outputs a discrete asset-state view (Bullish/Bearish/Neutral) together with a short list of key drivers.
    
    \item \textit{Technical Agent.} \addtextblue{Analyzes daily-price indicators such as moving averages, Moving Average Convergence Divergence (MACD), and Bollinger Bands~\cite{gencay1996non,wang2018predicting,day2023profitability}.} It produces a clear technical signal (Bullish/Bearish/Neutral) grounded in indicator-level evidence.
    
    \item \textit{News Agent.} \addtextblue{Monitors Bitcoin.com News.} It summarizes potentially market-moving events and estimates their likely directional impact with brief justification.
    
    \item \textit{Visual Agent.} \addtextblue{Receives daily OHLCV candlestick charts based on historical price data.} It inspects trend direction, salient candlestick patterns, and visually identifiable support/resistance levels.
\end{enumerate}
\subsection{Analyst Agent} 
After the four specialized data agents have collected and processed their respective information, the \textit{Analyst} synthesizes outputs from the \textit{Data Team} into high-level insights, identifying trends, anomalies, and opportunities, ensuring that downstream trading and risk decisions are grounded in coherent market understanding.

\subsubsection{Reflector Agent}
\label{sec:reflection}

The \textit{Reflector Agent} reviews each completed decision and turns its outcome into a lesson for later reuse. \addtextblue{Once the next opening price arrives, it compares the realized net portfolio return against the corresponding market return. The terminal opening price closes the final trajectory and writes its lesson to memory without generating a new trade.} What it records is not the raw trade but the judgment of whether the decision worked in its market context. This judgment makes the accumulated memory actionable rather than a mere log.

\subsubsection{Risk Manager Agent}

The \textit{Risk Manager} prioritizes capital preservation by qualitatively assessing market conditions. Instead of relying on rigid statistical metrics, it utilizes LLM to evaluate the insights from the \textit{Analyst Agent} alongside technical indicators. When high volatility or weak signals are detected, the agent issues risk warnings to the \textit{Trader Team}. \addtextblue{The resulting report is passed to the \textit{Trader Team} as contextual input for the final decision.}

\subsubsection{Memory Module}
\label{memory_module}

\addtextblue{The \textit{Memory Module} maintains a persistent repository of historical insights for reuse in later decisions.} \addtextblue{We instantiate it as a Financial ReasoningBank rather than a raw history buffer. For each completed trading trajectory, the system records the market query, analyst synthesis, risk assessment, retrieved context, trader proposals, executed action, realized P\&L, and market return. An LLM-as-a-judge labels the trajectory \textit{successful}, \textit{failed}, or \textit{neutral} from this financial feedback. A memory extractor distills it into compact, reusable trading principles rather than raw logs.} \addtextblue{A separate ReasoningBank is built for each run and asset, and relevant experiences are retrieved by vector similarity.}
\addtextblue{At inference time, the current market state is converted into a retrieval query, and the top-\(k\) relevant memories are injected into trader prompts. Every completed trajectory is judged at the next Open. If the LLM memory extractor returns no item, a deterministic fallback lesson is stored so the trajectory is not silently lost.}

\subsubsection{Trader Team}
\label{decision_module}

The \textit{Trader Team} is responsible for executing final trading decisions by integrating inputs from the \textit{Data Team}, \textit{Reflector}, \textit{Risk Manager} and external tools. It combines classical technical indicators (e.g., MACD, the Relative Strength Index, and mean-reversion strategies) with data-driven analyses to ensure robust and adaptive decision-making under volatile market conditions.

We instantiate three trader agents representing distinct trading styles: aggressive, conservative, and neutral. Each agent evaluates the same market context from its perspective: the aggressive agent prioritizes potential upside, the conservative agent emphasizes downside protection, and the neutral agent balances both considerations. \addtextblue{At each decision step $t$, each trader produces a continuous action \(a_{i,t} \in [-1,1]\): a positive value buys that fraction of remaining cash, a negative value sells that fraction of held asset, and zero leaves the allocation unchanged. Each action is accompanied by a short natural-language rationale. The final trading action is the weighted continuous aggregation of individual agent outputs.}
\addtextblue{Specifically, \(a_t = \operatorname{clip}(\sum_{i=1}^{3} w_{i,t} \cdot a_{i,t}, -1, 1)\), subject to \(\sum_{i=1}^{3} w_{i,t} = 1\). Here, $w_{i,t}$ denotes the normalized weight of agent $i$ at time $t$, and $a_{i,t}$ is its continuous trader action.}

\paragraph{Dynamic Weight Allocation via Reinforcement Learning.}  

To handle non-stationary market dynamics, we formulate trader weight assignment as a continuous-control RL problem, which we refer to as \textit{Market-Feedback-Alignment}. Concretely, we model the refinement process as an episodic Markov decision process  \((\mathcal{S}, \mathcal{A}, \mathcal{P}, \mathcal{R})\), where the policy does not change the underlying trader agents or their prompts, but learns how to allocate their contributions using market feedback.

\textbf{State.} \addtextblue{At each trading step \(t\), the state \(s_t \in \mathcal{S}\) summarizes both market context and trader-level credit information: (i) recent price/return history and technical indicators, (ii) the previous-step trader weights \(\mathbf{w}_{t-1} \in \Delta^2\), (iii) the candidate actions \(\{a_{i,t}\}_{i=1}^{3}\) proposed by the three style-specific trader agents, collected as \(\mathbf{a}_t=[a_{1,t},a_{2,t},a_{3,t}]\), and (iv) a counterfactual contribution vector \(\mathbf{C}_{t}=[C_{t}^{\mathrm{aggr}}, C_{t}^{\mathrm{cons}}, C_{t}^{\mathrm{neut}}]\).}

\addtextblue{Counterfactual Contribution Memory estimates each trader's recent standalone reliability under the actually realized market sequence. Let \(\rho_{t+1}=(p_{t+1}-p_t)/p_t\) denote the next-step market return after trader \(i\) proposes action \(a_{i,t}\).} \addtextblue{With the continuous action range \(a_{i,t}\in[-1,1]\) under the long-only trading protocol, the hypothetical contribution is}
\begin{equation}
\addtextblue{h_{i,t} =
\begin{cases}
a_{i,t}\rho_{t+1}, & a_{i,t}>0,\\
0, & a_{i,t}=0,\\
\lvert a_{i,t}\rvert\max(-\rho_{t+1},0), & a_{i,t}<0,
\end{cases}}
\end{equation}
\addtextblue{A negative action is interpreted as moving to cash and receives positive credit only when its magnitude avoids a subsequent down move. The contribution memory is \(C_{i,t}=\frac{1}{K}\sum_{\tau=t-K}^{t-1} h_{i,\tau}\).}
\addtextblue{Here, \(K\) is the contribution-memory window, distinct from the retrieval limit \(k\). The denominator remains \(K\) during warm-up, so missing pre-history terms are equivalent to zeros.}
\addtextblue{Thus the gating state is \(s_t=[\mathbf{w}_{t-1}, \mathrm{MarketFeatures}_t, \mathbf{a}_t, \mathbf{C}_t]\): PPO receives the net portfolio-return reward defined below and numerical CCM, whereas ReasoningBank memories affect PPO indirectly through the trader actions generated from their prompts.}

\textbf{Action.} The action is a continuous weight vector $\mathbf{w}_t \in \mathcal{A}$ that lies on the probability simplex,
\(\mathbf{w}_t \in \Delta^2, w_{i,t} \ge 0,\ \sum_{i=1}^{3} w_{i,t} = 1, \)
which modulates each trader agent's influence on the aggregated trading action.

\textbf{Reward (market feedback).} After executing the aggregated action, the environment returns a scalar reward $r_t = \mathcal{R}(s_t, \mathbf{w}_t)$ derived from realized trading performance. \addtextblue{In practice, we use realized returns net of transaction costs as the primary feedback signal.}
\addtextblue{The first validation decision only initializes a pending state-action record and stores no transition. Each later Open settles the previous pending transition, and the final Open settles the last one, so every stored transition has a realized next-Open reward.}

\section{Experiment}
We evaluate Agent4Crypto through four research questions: (1)  \textbf{Research Question 1 (RQ1):} How does it compare with state-of-the-art traditional and LLM-based cryptocurrency trading models?  (2) \textbf{Research Question 2 (RQ2):} How much does each component contribute to overall performance? (3)  \textbf{Research Question 3 (RQ3):} How do weight refinement strategies (direct vs. RL-based) affect trader weights and trading performance? (4) \textbf{Research Question 4 (RQ4):} What explains Agent4Crypto's risk-avoidance advantage over CryptoTrade in a matched decision?
\subsection{Experimental Setup}
\subsubsection{Datasets}
\label{datasets}
\addtextblue{We evaluate on BTC, ETH, and XRP. Their test windows collectively span bullish, neutral/sideways, and bearish conditions.} \addtextblue{Each asset is split chronologically into a validation window followed by a test window. BTC uses validation 2026-03-14--2026-04-13 and test 2026-04-13--2026-06-27. ETH uses validation 2026-03-17--2026-04-16 and test 2026-04-16--2026-06-30. XRP uses validation 2026-04-06--2026-05-06 and test 2026-05-06--2026-07-20. All validation and test windows begin after GPT-5.4 was released on 2026-03-05.}
\addtextblue{For Agent4Crypto, the validation trajectory is replayed once to train the PPO gating policy. Each action is settled at the following Open and its fee-adjusted net portfolio return is used as the reward. At the validation boundary, the remaining update is applied and the policy is frozen. The portfolio, memory, contribution history, and pending transitions are reset before test replay. Only the recent Open-price history needed for lookback features carries over. For tunable baselines, validation performance selects the parameter or lookback configuration, which is then fixed.}

\addtextblue{To prevent temporal (look-ahead) leakage in our experiments, every prompt-facing input is constructed in a strict point-in-time manner using only information available by the UTC Open on decision day \(t\). Daily market candles are loaded from the Coinbase records in the release, and the current row's High, Low, Close, and Volume are excluded from the prompt. Technical features use only Open prices available no later than the decision time, while the chart contains only rows strictly before \(t\) with a 30-observation lookback. Dated on-chain observations are read from the Coin Metrics Community API snapshots at a one-day lag. News is read from the one-day-lagged archive and retained only when its publication timestamp, normalized to UTC, precedes the decision timestamp. Thus, no value, chart row, on-chain observation, or news item that first becomes available after a decision is exposed to that decision.}

\setcounter{table}{1}
\begin{table*}[t]
\centering
\caption{\addtextblue{Performance comparison across BTC, ETH, and XRP. All baselines and Agent4Crypto results are averaged over five runs with different random seeds or repeated LLM calls. The best value is bold and the second-best is underlined.}}
\label{tab:three-market-baselines}
\small
\setlength{\tabcolsep}{4pt}
\begin{tabular}{l|ccc|ccc|ccc}
\toprule
\addtextblue{\textbf{Strategy}} & \multicolumn{3}{c|}{\addtextblue{\textbf{Total Return (\%) $\uparrow$}}} & \multicolumn{3}{c|}{\addtextblue{\textbf{Daily Return (\%) $\uparrow$}}} & \multicolumn{3}{c}{\addtextblue{\textbf{Sharpe Ratio $\uparrow$}}} \\
\cmidrule(lr){2-4} \cmidrule(lr){5-7} \cmidrule(lr){8-10}
 & \addtextblue{\textbf{BTC}} & \addtextblue{\textbf{ETH}} & \addtextblue{\textbf{XRP}} & \addtextblue{\textbf{BTC}} & \addtextblue{\textbf{ETH}} & \addtextblue{\textbf{XRP}} & \addtextblue{\textbf{BTC}} & \addtextblue{\textbf{ETH}} & \addtextblue{\textbf{XRP}} \\
\midrule
\addtextblue{Buy and Hold} & \addtextblue{-15.24} & \addtextblue{-31.81} & \addtextblue{-22.37} & \addtextblue{\ms{-0.20}{1.93}} & \addtextblue{\ms{-0.48}{2.47}} & \addtextblue{\ms{-0.31}{2.35}} & \addtextblue{-0.10} & \addtextblue{-0.19} & \addtextblue{-0.13} \\
\addtextblue{SMA} & \addtextblue{1.19} & \addtextblue{\underline{-10.78}} & \addtextblue{-11.39} & \addtextblue{\ms{0.02}{1.03}} & \addtextblue{\ms{\underline{-0.15}}{0.99}} & \addtextblue{\ms{-0.15}{1.29}} & \addtextblue{0.02} & \addtextblue{-0.15} & \addtextblue{-0.12} \\
\addtextblue{SLMA} & \addtextblue{\underline{3.78}} & \addtextblue{-16.51} & \addtextblue{-11.16} & \addtextblue{\ms{\underline{0.06}}{1.13}} & \addtextblue{\ms{-0.23}{1.19}} & \addtextblue{\ms{-0.15}{1.16}} & \addtextblue{\underline{0.05}} & \addtextblue{-0.20} & \addtextblue{-0.13} \\
\addtextblue{MACD} & \addtextblue{-2.41} & \addtextblue{-11.51} & \addtextblue{-15.92} & \addtextblue{\ms{-0.03}{1.13}} & \addtextblue{\ms{-0.16}{1.11}} & \addtextblue{\ms{-0.22}{1.35}} & \addtextblue{-0.02} & \addtextblue{-0.14} & \addtextblue{-0.16} \\
\addtextblue{Bollinger Bands} & \addtextblue{-11.48} & \addtextblue{-28.89} & \addtextblue{-11.61} & \addtextblue{\ms{-0.15}{1.43}} & \addtextblue{\ms{-0.43}{2.25}} & \addtextblue{\ms{-0.15}{1.96}} & \addtextblue{-0.11} & \addtextblue{-0.19} & \addtextblue{\underline{-0.07}} \\
\midrule
\addtextblue{LSTM} & \addtextblue{-17.34} & \addtextblue{-16.67} & \addtextblue{-20.51} & \addtextblue{\ms{-0.24}{1.63}} & \addtextblue{\ms{-0.23}{1.76}} & \addtextblue{\ms{-0.28}{2.00}} & \addtextblue{-0.15} & \addtextblue{\underline{-0.12}} & \addtextblue{-0.16} \\
\addtextblue{Informer} & \addtextblue{-10.83} & \addtextblue{-19.42} & \addtextblue{-11.25} & \addtextblue{\ms{-0.15}{1.15}} & \addtextblue{\ms{-0.28}{1.32}} & \addtextblue{\ms{-0.15}{1.17}} & \addtextblue{-0.13} & \addtextblue{-0.21} & \addtextblue{-0.14} \\
\addtextblue{AutoFormer} & \addtextblue{-7.65} & \addtextblue{-20.57} & \addtextblue{-10.89} & \addtextblue{\ms{-0.10}{1.03}} & \addtextblue{\ms{-0.30}{1.47}} & \addtextblue{\ms{-0.15}{1.26}} & \addtextblue{-0.10} & \addtextblue{-0.20} & \addtextblue{-0.11} \\
\addtextblue{TimesNet} & \addtextblue{-6.26} & \addtextblue{-14.29} & \addtextblue{-11.72} & \addtextblue{\ms{-0.08}{0.96}} & \addtextblue{\ms{-0.20}{1.06}} & \addtextblue{\ms{-0.16}{1.10}} & \addtextblue{-0.09} & \addtextblue{-0.18} & \addtextblue{-0.15} \\
\addtextblue{PatchTST} & \addtextblue{-13.04} & \addtextblue{-15.36} & \addtextblue{\underline{-5.78}} & \addtextblue{\ms{-0.18}{1.37}} & \addtextblue{\ms{-0.21}{1.35}} & \addtextblue{\ms{\underline{-0.07}}{0.94}} & \addtextblue{-0.13} & \addtextblue{-0.16} & \addtextblue{-0.10} \\
\midrule
\addtextblue{DQN} & \addtextblue{-7.74} & \addtextblue{-22.23} & \addtextblue{-19.34} & \addtextblue{\ms{-0.10}{1.14}} & \addtextblue{\ms{-0.33}{1.77}} & \addtextblue{\ms{-0.27}{1.59}} & \addtextblue{-0.05} & \addtextblue{-0.15} & \addtextblue{-0.17} \\
\addtextblue{PPO} & \addtextblue{-5.75} & \addtextblue{-25.45} & \addtextblue{-11.18} & \addtextblue{\ms{-0.08}{0.84}} & \addtextblue{\ms{-0.38}{1.97}} & \addtextblue{\ms{-0.15}{1.13}} & \addtextblue{0.02} & \addtextblue{-0.16} & \addtextblue{-0.11} \\
\addtextblue{SAC} & \addtextblue{-10.83} & \addtextblue{-18.40} & \addtextblue{-14.33} & \addtextblue{\ms{-0.15}{1.25}} & \addtextblue{\ms{-0.27}{1.20}} & \addtextblue{\ms{-0.20}{1.54}} & \addtextblue{-0.12} & \addtextblue{-0.22} & \addtextblue{-0.13} \\
\midrule
\addtextblue{CryptoTrade} & \addtextblue{-5.12} & \addtextblue{-16.29} & \addtextblue{-16.08} & \addtextblue{\ms{-0.06}{1.11}} & \addtextblue{\ms{-0.23}{0.98}} & \addtextblue{\ms{-0.23}{1.18}} & \addtextblue{-0.06} & \addtextblue{-0.24} & \addtextblue{-0.19} \\
\addtextblue{Agent4Crypto} & \addtextblue{\textbf{6.66}} & \addtextblue{\textbf{-4.77}} & \addtextblue{\textbf{-2.36}} & \addtextblue{\ms{\textbf{0.09}}{0.89}} & \addtextblue{\ms{\textbf{-0.06}}{0.74}} & \addtextblue{\ms{\textbf{-0.02}}{0.48}} & \addtextblue{\textbf{0.10}} & \addtextblue{\textbf{-0.08}} & \addtextblue{\textbf{-0.04}} \\
\bottomrule
\end{tabular}%
\end{table*}
\FloatBarrier

\subsubsection{Evaluation Metrics}
\label{evaluation_metrics}
We evaluate Agent4Crypto against baselines using multiple widely-adopted financial performance metrics \cite{li2024cryptotrade}:
\addtextblue{\textbf{Total Return:} Measures the overall profit or loss over the trading period by comparing the final net capital with the initial net capital. \textbf{Daily Return Mean ($\bar{r}$):} the arithmetic mean of daily returns over the trading horizon, capturing the average day-to-day performance. \textbf{Daily Return Standard Deviation ($\sigma$):} indicates the volatility and risk associated with the daily performance of the trading strategy. \textbf{Sharpe Ratio:} a risk-adjusted performance metric defined as \(\text{Sharpe Ratio} = \frac{\bar{r} - r_f}{\sigma}\), where $r_f = 0$ in our experiments \cite{cheng2024sociodojo, li2024cryptotrade}. It balances profitability and risk, and is widely used to evaluate trading strategies.}

\subsubsection{Baseline Methods}

\addtextblue{We compare Agent4Crypto against rule-based, ML\&DL, RL, and LLM-based baselines, all evaluated in the same environment. The rule-based baselines are Buy-and-Hold, Simple Moving Average (SMA) and Short/Long Moving Average Crossover (SLMA) \cite{gencay1996non}, Moving Average Convergence Divergence (MACD) \cite{wang2018predicting}, and Bollinger Bands \cite{day2023profitability}. SLMA takes a long position when the short-window average exceeds the long-window one and a short position otherwise. The ML\&DL baselines are LSTM, Informer, AutoFormer, TimesNet, and PatchTST. The RL baselines are DQN, PPO, and SAC. CryptoTrade is the LLM-based baseline.}

\subsubsection{Execution Protocol and Costs}
\label{execution_protocol}
\addtextblue{All strategies share the same execution convention and continuous position action space. Each run starts with \$1,000,000, split evenly between cash and the asset. On decision day \(t\), the action \(a_t\in[-1,1]\) is clipped to this interval: \(a_t>0\) buys \(a_t\) times the remaining cash, \(a_t<0\) sells \(\lvert a_t\rvert\) times the held units, and \(a_t=0\) leaves the allocation unchanged. Trades execute at the current Open, portfolio wealth is valued at the next Open, and a \(0.1\%\) fee is charged on each traded notional.}

\addtextblue{To keep the comparison robust to randomness in training and in LLM sampling, all baselines are evaluated with five independent random seeds, 42, 43, 44, 45, and 46. Every LLM-based method is additionally run five times with repeated LLM generations. The reported metrics are averaged over these runs.}

\addtextblue{All LLM calls in Agent4Crypto and CryptoTrade use GPT-5.4 with a 4096-token output cap and a default temperature of 0.0. Agent4Crypto uses a 90-day market lookback, a 30-observation candlestick chart with volume and 5, 10, and 20-period moving averages. The retrieval limit is \(k=3\), and the contribution-memory window is \(K=5\). The PPO gate is a two-layer multilayer perceptron with 64 Tanh units per layer, a Dirichlet concentration factor of 10 with a 0.1 offset during validation, a 50-step rollout horizon, discount factor 0.99, learning rate \(5\times10^{-4}\), and 10 update epochs. Malformed trader outputs are mapped to a neutral action, and an empty memory extraction triggers a deterministic fallback lesson. Deep baselines use a 30-step lookback, batch size 32, and 10 epochs with Adam learning rates of \(10^{-3}\), except for Informer, which uses \(10^{-4}\). RL baselines use 50 episodes, a 10-step lookback, 64 hidden units, batch size 32, learning rate \(3\times10^{-4}\), and discount factor 0.99.}

\addtextblue{For RQ3, we compare three trader-weight variants under the same protocol. The EW (equal weight) variant keeps equal weights. The DR (direct refinement) variant applies direct market feedback. Agent4Crypto learns the weights with PPO and Counterfactual Contribution Memory. They use the same traders, actions, fee, and chronological validation/test periods. Only the weight-refinement mechanism differs.}

\addtextblue{Agent4Crypto-DR is the Direct Refinement variant. It starts from uniform weights. At the start of decision day $t$, after observing the new price \(p_t\) and before the trader actions are generated, it computes the realized move \(g_t=(p_t-p_{t-1})/p_{t-1}\). For trader $i$, the weight increment is \(\delta_{i,t}=0\) when \(\lvert g_t\rvert\leq 0.1\%\), and otherwise \(\delta_{i,t}=\alpha\max(\lvert a_{i,t-1}\rvert,0.01)\operatorname{sign}(a_{i,t-1})\operatorname{sign}(g_t)\) with $\alpha=0.01$. It then sets \(w_{i,t}=\max(w_{i,t-1}+\delta_{i,t},0.01)\) for the Aggressive, Conservative, and Neutral traders and renormalizes the three weights to sum to one. A nonzero action aligned with the realized move therefore gains weight, an opposing action loses weight, and a zero action is unchanged.}

\section{Experimental Results}
We evaluate Agent4Crypto through performance comparisons with baselines, ablation studies, adaptive weight analysis, and a case study of risk avoidance against CryptoTrade.
\begin{table*}[tb]
    \setcounter{table}{2}
    \centering
    \caption{\addtextblue{High-level module ablation study of Agent4Crypto across BTC, ETH, and XRP. Each variant removes one major architectural capability while retaining the remaining system. All baselines and Agent4Crypto results are averaged over five runs with different random seeds or repeated LLM calls. The best value is bold and the second-best is underlined.}}
    \small
\begin{tabular}{l|ccc|ccc|ccc}
\toprule
\addtextblue{\textbf{Variant}} & \multicolumn{3}{c|}{\addtextblue{\textbf{Total Return (\%) $\uparrow$}}} & \multicolumn{3}{c|}{\addtextblue{\textbf{Daily Return (\%) $\uparrow$}}} & \multicolumn{3}{c}{\addtextblue{\textbf{Sharpe Ratio $\uparrow$}}} \\
\cmidrule(lr){2-4} \cmidrule(lr){5-7} \cmidrule(lr){8-10}
 & \addtextblue{\textbf{BTC}} & \addtextblue{\textbf{ETH}} & \addtextblue{\textbf{XRP}} & \addtextblue{\textbf{BTC}} & \addtextblue{\textbf{ETH}} & \addtextblue{\textbf{XRP}} & \addtextblue{\textbf{BTC}} & \addtextblue{\textbf{ETH}} & \addtextblue{\textbf{XRP}} \\
\midrule
\addtextblue{w/o Data Team Specialization} & \addtextblue{3.87} & \addtextblue{\underline{-4.83}} & \addtextblue{-2.57} & \addtextblue{\ms{0.05}{0.56}} & \addtextblue{\ms{\underline{-0.06}}{0.49}} & \addtextblue{\ms{-0.03}{0.31}} & \addtextblue{0.09} & \addtextblue{-0.12} & \addtextblue{-0.10} \\
\addtextblue{w/o Inter-team Collaboration} & \addtextblue{6.23} & \addtextblue{-5.60} & \addtextblue{-6.25} & \addtextblue{\ms{0.08}{1.01}} & \addtextblue{\ms{-0.07}{0.81}} & \addtextblue{\ms{-0.08}{0.66}} & \addtextblue{0.08} & \addtextblue{\underline{-0.09}} & \addtextblue{-0.12} \\
\addtextblue{w/o Trader Competition} & \addtextblue{\underline{6.63}} & \addtextblue{-5.35} & \addtextblue{\underline{-2.41}} & \addtextblue{\ms{\underline{0.09}}{0.84}} & \addtextblue{\ms{-0.07}{0.75}} & \addtextblue{\ms{\underline{-0.02}}{0.43}} & \addtextblue{\textbf{0.10}} & \addtextblue{-0.09} & \addtextblue{\underline{-0.06}} \\
\midrule
\addtextblue{Agent4Crypto} & \addtextblue{\textbf{6.66}} & \addtextblue{\textbf{-4.77}} & \addtextblue{\textbf{-2.36}} & \addtextblue{\ms{\textbf{0.09}}{0.89}} & \addtextblue{\ms{\textbf{-0.06}}{0.74}} & \addtextblue{\ms{\textbf{-0.02}}{0.48}} & \addtextblue{\underline{0.10}} & \addtextblue{\textbf{-0.08}} & \addtextblue{\textbf{-0.04}} \\
\bottomrule
\end{tabular}%
\label{tab:ablation_full}
\end{table*}
\subsection{Performance Comparison (RQ1)}
\addtextblue{The performance comparison in Table~\ref{tab:three-market-baselines} shows that Agent4Crypto achieves the best total return on all three assets, reaching 6.66\% on BTC, $-4.77\%$ on ETH, and $-2.36\%$ on XRP. Among the baselines, SLMA on BTC (3.78\%), SMA on ETH ($-10.78\%$), and PatchTST on XRP ($-5.78\%$) have the strongest total returns. Agent4Crypto improves on them by 2.88, 5.30, and 3.42 percentage points. It also has the best Sharpe ratio on every asset, above SLMA on BTC (0.05), LSTM on ETH ($-0.12$), and Bollinger Bands on XRP ($-0.07$). All rule-based, ML\&DL, and RL baselines, as well as CryptoTrade, underperform Agent4Crypto.}

\subsection{Ablation Study (RQ2)}
\addtextblue{To assess each architectural component, we perform individual ablations of Agent4Crypto: Data Team specialization, inter-team collaboration, and Trader competition (Table~\ref{tab:ablation_full}). Each variant removes one module while keeping the rest unchanged.}
\paragraph{Effect of Data Team Specialization.}
\addtextblue{This ablation bypasses all four specialized Data Agents while retaining the same textual, numerical, and chart evidence for direct synthesis by the Analyst, isolating the value of specialized multimodal interpretation.}
\paragraph{Effect of Inter-team Collaborative Reasoning.}
\addtextblue{We jointly remove the Reflector, Financial ReasoningBank, and Risk Manager while preserving the Data Team, Analyst, Trader Team, and trading environment. This evaluates retrospective feedback, retrieved experience, and qualitative risk context as a single cross-role collaboration capability.}
\paragraph{Effect of Competitive Trader Dynamics.}
\addtextblue{We retain only the Neutral Trader and disable both PPO gating and CCM. Comparing this variant with the full system directly tests whether heterogeneous trader proposals and adaptive, contribution-aware aggregation add value beyond a single balanced trader.}
\addtextblue{The ablations establish that the complete system is the only configuration that stays at the top on every asset. No single ablated variant dominates. Disabling trader competition is nearly as good on BTC (6.63\% vs.\ 6.66\%) but falls behind on XRP ($-2.41\%$ vs.\ $-2.36\%$). Removing Data Team specialization is marginally better on ETH alone ($-4.83\%$ vs.\ $-4.77\%$) but costs 2.79 points on BTC and 0.21 on XRP. Each component earns its place on a different asset, which is why dropping any one of them hurts somewhere. The clearest evidence is inter-team collaboration. Without it, total return on XRP drops from $-2.36\%$ to $-6.25\%$ and the Sharpe ratio from $-0.04$ to $-0.12$. The full system combines all three capabilities and thereby avoids the failure mode that appears whenever one is removed. It is best or matches the best result on all three assets, whereas every ablation is worse on at least one.}
\subsection{Influence of Different Trader Weights (RQ3)}
\addtextblue{Table~\ref{tab:weight-variants} compares three weighting rules under the same traders, actions, fees, and splits. The variants differ only in how they set the weights: Agent4Crypto-EW keeps them equal, Agent4Crypto-DR applies Direct Refinement based on the realized market direction, and Agent4Crypto learns them with PPO and Counterfactual Contribution Memory.}

\begin{table*}[t]
\centering
\caption{\addtextblue{Influence of different trader weights. All baselines and Agent4Crypto results are averaged over five runs with different random seeds or repeated LLM calls. The best value is bold and the second-best is underlined.}}
\label{tab:weight-variants}
\small
\begin{tabular}{l|ccc|ccc|ccc}
\toprule
\addtextblue{\textbf{Variant}} & \multicolumn{3}{c|}{\addtextblue{\textbf{Total Return (\%) $\uparrow$}}} & \multicolumn{3}{c|}{\addtextblue{\textbf{Daily Return (\%) $\uparrow$}}} & \multicolumn{3}{c}{\addtextblue{\textbf{Sharpe Ratio $\uparrow$}}} \\
\cmidrule(lr){2-4} \cmidrule(lr){5-7} \cmidrule(lr){8-10}
 & \addtextblue{\textbf{BTC}} & \addtextblue{\textbf{ETH}} & \addtextblue{\textbf{XRP}} & \addtextblue{\textbf{BTC}} & \addtextblue{\textbf{ETH}} & \addtextblue{\textbf{XRP}} & \addtextblue{\textbf{BTC}} & \addtextblue{\textbf{ETH}} & \addtextblue{\textbf{XRP}} \\
\midrule
\addtextblue{Agent4Crypto-EW} & \addtextblue{\underline{6.13}} & \addtextblue{-5.22} & \addtextblue{-4.11} & \addtextblue{\ms{\underline{0.08}}{0.99}} & \addtextblue{\ms{-0.07}{0.73}} & \addtextblue{\ms{-0.05}{0.50}} & \addtextblue{\underline{0.08}} & \addtextblue{-0.09} & \addtextblue{-0.11} \\
\addtextblue{Agent4Crypto-DR} & \addtextblue{5.82} & \addtextblue{\underline{-5.12}} & \addtextblue{\underline{-2.85}} & \addtextblue{\ms{0.08}{0.95}} & \addtextblue{\ms{\underline{-0.07}}{0.76}} & \addtextblue{\ms{\underline{-0.04}}{0.47}} & \addtextblue{0.08} & \addtextblue{\underline{-0.09}} & \addtextblue{\underline{-0.08}} \\
\addtextblue{Agent4Crypto} & \addtextblue{\textbf{6.66}} & \addtextblue{\textbf{-4.77}} & \addtextblue{\textbf{-2.36}} & \addtextblue{\ms{\textbf{0.09}}{0.89}} & \addtextblue{\ms{\textbf{-0.06}}{0.74}} & \addtextblue{\ms{\textbf{-0.02}}{0.48}} & \addtextblue{\textbf{0.10}} & \addtextblue{\textbf{-0.08}} & \addtextblue{\textbf{-0.04}} \\
\bottomrule
\end{tabular}%
\end{table*}
\FloatBarrier
\addtextblue{The weighting mechanism matters: Agent4Crypto is the only variant that is best on all three assets and on all three metrics, and the ordering of the variants reveals why. The equal-weight reference lags most on XRP ($-4.11\%$), where averaging a bullish, a bearish, and a neutral view dilutes the one style the market actually rewarded. The Direct Refinement variant does better ($-2.85\%$) but still trails Agent4Crypto ($-2.36\%$). In magnitude the gains over Agent4Crypto-EW and Agent4Crypto-DR are 0.53 and 0.84 points on BTC, 0.45 and 0.35 on ETH, and 1.75 and 0.49 on XRP. Equal weighting never adapts. Direct Refinement adapts too narrowly. Only Agent4Crypto reallocates influence from longer-horizon feedback about which trader's style fit the recent market.}
\begin{figure*}[t]
    \centering
    \includegraphics[width=0.92\textwidth]{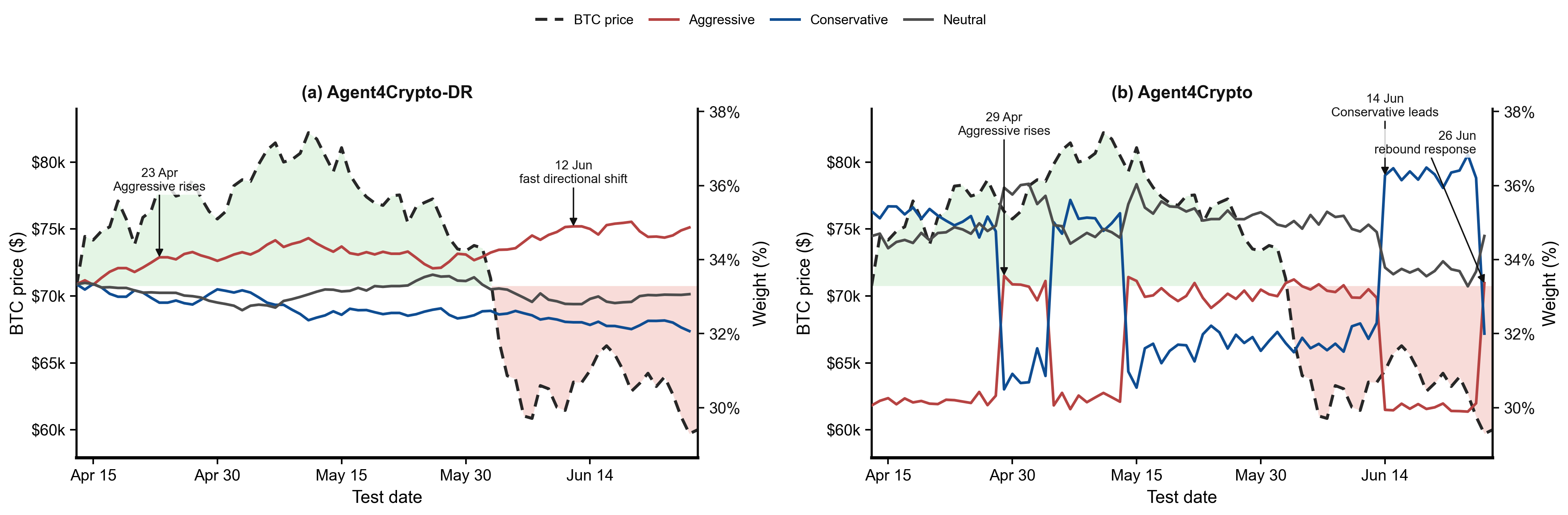}
    \Description{\addtextblue{Two side-by-side BTC panels with one shared legend. Each panel includes both vertical scales. The left vertical scale reports BTC opening price in U.S. dollars, while the right vertical scale reports aggressive, conservative, and neutral trader weights as percentages. The pale green and pale red areas show price movement above and below the first observed price and do not denote uncertainty. Panel (a) shows Agent4Crypto-DR and panel (b) shows Agent4Crypto, with arrows marking the main observed weight reallocations.}}
    \caption{\addtextblue{BTC case study: trader-weight adaptation under Agent4Crypto-DR (a) and Agent4Crypto (b). BTC price is read from the left vertical axis in U.S. dollars and trader weights from the right vertical axis as percentage shares.}}
    \label{fig:btc_weight_case}
\end{figure*}

\addtextblue{The advantage of the full model over the Direct Refinement variant is visible in how the two move trader weight through successive market phases. Figure~\ref{fig:btc_weight_case} overlays the BTC trajectory with the weight paths of Agent4Crypto-DR and Agent4Crypto, where the difference is not the direction they eventually take but when and how smoothly they take it.}

\addtextblue{In the early test segment, including the fluctuations before 1 May, Agent4Crypto-DR reacts more immediately to realized direction and increases the Aggressive weight during the upward BTC movement. Agent4Crypto is less abrupt, which is consistent with a policy that incorporates longer-horizon feedback rather than only the latest price change. During the local weakness around 1 May, Agent4Crypto increases the Neutral weight, preserving a more balanced response while the market signal is mixed.}

\addtextblue{Around the decline beginning near 15 May, Agent4Crypto gradually shifts influence away from Aggressive, with the larger reallocation going to Conservative while retaining a substantial Neutral allocation. The separation becomes most consequential after 1 June, when BTC enters a sustained drawdown: by mid-June, the full model assigns the largest share to Conservative and thereby reduces exposure to the subsequent selloff. Agent4Crypto-DR remains closer to its recent Direct Refinement update and does not exhibit the same longer-horizon defensive reallocation.}

\addtextblue{At the final decision date, 26 June, the next Open moves upward on 27 June. Agent4Crypto responds by raising Aggressive and Neutral together and lowering Conservative, matching the short rebound without abandoning the defensive allocation accumulated during the preceding decline. This BTC example highlights the advantage of Agent4Crypto across successive market phases. Its memory-aware policy preserves defensive allocation through the drawdown while still responding to the rebound, whereas Agent4Crypto-DR follows the latest realized direction and does not retain the same cross-phase adjustment.}

\subsection{Case Study: Risk Avoidance (RQ4)}
\addtextblue{To further illustrate the decision process, Table~\ref{tab:case_study} traces a single ETH decision through evidence decomposition, inter-team risk reasoning, competitive trader proposals, and PPO/CCM aggregation to show how the modules produce the correct action.}
\li{The matched decision is dated 2026-06-23 UTC. Both systems saw only information available up to that point. ETH then fell from \$1,726.12 to \$1,665.04 by the next open, a market return of $-3.5386\%$.}

\li{\textbf{Context:} The two systems faced the same contradictory evidence. On-chain fundamentals pointed mildly upward. They included 2.84 million transactions, 817 thousand active addresses, and roughly 151 ETH in fees. Every short-horizon signal pointed down. These signals included MACD at $-65.36$, risk-off exchange-traded fund (ETF) flows, a Taiko bridge exploit adding security stress, and a large leveraged short opened across BTC, SOL, and ETH. The visual report placed ETH in bearish-to-sideways consolidation below the \$1,750--1,800 resistance zone. The decision therefore hinged on which of these conflicting signals each system trusted.}

\begin{table*}[t]
\centering
\caption{Module-level case study: ETH decision on 23 June 2026. Specialized evidence decomposition, inter-team risk reasoning, competitive traders, and RL aggregation produced the correct action that CryptoTrade's single-agent pipeline did not.}
\label{tab:case_study}
\small
\setlength{\tabcolsep}{4pt}
\begin{tabular}{@{}p{0.18\textwidth}|p{0.37\textwidth}|p{0.37\textwidth}@{}}
\toprule
\textbf{Module} & \multicolumn{1}{>{\centering\arraybackslash}p{0.37\textwidth}|}{\textbf{Agent4Crypto}} & \multicolumn{1}{>{\centering\arraybackslash}p{0.37\textwidth}}{\textbf{CryptoTrade}} \\
\midrule
\textbf{Data Team specialization} &
Four independent reports separated healthy fundamentals from bearish technical momentum, risk-off news, and a visual consolidation below \$1,750--1,800. This prevented a one-day indicator flip from dominating the evidence. &
	The baseline used one GPT-5.4 pipeline with fused on-chain/technical, news, and reflection prompts. It had no specialized Asset, Technical, News, and Visual agents. Its synthesis foregrounded ``MACD, SMA, and SLMA flipped to buy.'' \\
\midrule
\textbf{Inter-team risk} &
The Analyst synthesized a neutral-to-bearish short-term trend. The separate Risk Manager issued ``Caution / Reduce Exposure,'' and the retrieved ReasoningBank included the guardrail ``oversold readings require de-risking, not automatic reversal buying.'' &
There was no separate Risk Manager or Financial ReasoningBank to turn the mixed reports into an explicit risk constraint. The single trader acknowledged ETF outflows and short positioning but still formed a ``modest net-long bias.'' \\
\midrule
\textbf{Trader competition} &
Aggressive, Conservative, and Neutral traders independently proposed $-0.35$, $-0.25$, and $-0.25$. Their agreement made the defensive direction robust to any one trader's mistake. &
	One GPT-5.4 trader produced a single $+0.37$ action. There was no heterogeneous proposal set whose disagreement or agreement could be inspected. \\
\midrule
\textbf{RL/CCM aggregation and outcome} &
PPO/Counterfactual Contribution Memory assigned weights $(0.293,0.360,0.347)$ to Aggressive/Conservative/Neutral, yielding $-0.2793$. The resulting position was only $3.27\times10^{-8}$ ETH, and wealth stayed at \$952,265.82. &
The single-agent action increased the invested fraction from 62.07\% to 75.45\%. Wealth fell from \$883,947.13 to \$860,018.41, a daily return of $-2.7070\%$. \\
\bottomrule
\end{tabular}
\end{table*}

\li{\textbf{Mechanism-level interpretation:} The two systems diverged because only Agent4Crypto had a chain of modules that could act on the contradiction. Its Data Agents kept the bearish technical and visual evidence visible beside the healthy fundamentals instead of letting one signal dominate. The Analyst, Risk Manager, and ReasoningBank then converted that conflict into an explicit reduce-exposure instruction. The three traders independently reached the same sell direction. The PPO/CCM gate weighted the Conservative and Neutral proposals most heavily. CryptoTrade's single trader read the short-horizon indicator flip as a buy and raised the invested fraction to 75.45\%. The day's 2.7070-point return gap therefore traces directly to these modules.}

\section{Conclusion}
\label{sec:conclusion}
We introduce Agent4Crypto, a multi-agent LLM-based framework for cryptocurrency trading that integrates structured inter- and intra-team collaboration with competitive trader dynamics, aligned with real-time market feedback. By explicitly modeling diverse trader styles (Aggressive, Neutral, Conservative) and adaptively adjusting their influence via reinforcement learning, the system generates robust, informed trading decisions and adapts to evolving market regimes. Extensive experiments \addtextblue{on BTC, ETH, and XRP} show that Agent4Crypto exceeds the best evaluated baseline on all three assets and also exceeds CryptoTrade on all three assets. \addtextblue{Ablation results show that removing any single capability, namely Data Team specialization, inter-team collaboration, or trader competition, degrades performance on at least one asset. The effect is clearest for inter-team collaboration on XRP.} A case study further shows why Agent4Crypto reduced exposure before a subsequent decline: specialized Data Agents, inter-team risk reasoning, trader agreement, and PPO/CCM aggregation reconciled conflicting signals, whereas CryptoTrade interpreted the same evidence as a buy signal and increased exposure. These results highlight the potential of multi-agent LLM architectures with adaptive, market-aligned feedback for robust, interpretable, and high-performing AI-driven cryptocurrency trading.

\section{Ethical Considerations}
Agent4Crypto is intended solely as a research framework for analyzing cryptocurrency markets and should not be treated as financial advice or deployed without independent oversight. Cryptocurrency trading is highly volatile, and an apparently effective strategy can generate substantial losses when market conditions change, data are delayed or manipulated, or the model behaves unpredictably. The framework may also be misused for automated speculative trading at scale, which could amplify market stress or expose inexperienced users to losses. To mitigate these risks, any real-world deployment should retain human supervision, enforce exposure and loss limits, use authenticated and rate-limited data sources, clearly disclose uncertainty and system limitations, and undergo monitoring and stress testing before use.

\printbibliography
\end{document}